# An Efficient Security Model for Industrial Internet of Things (IIoT) System Based on Machine Learning Principles

**Sahar L. Qaddoori***  
sahar.qaddoori@uoninevah.edu.iq

**Qutaiba I. Ali****  
qut1974@gmail.com

* Electronic Engineering Departement, Collage of Electronics Engineering, Ninevah University, Mosul, Iraq  
** Computer Engineering Department, Collage of Engineering, University of Mosul, Mosul, Iraq

*ABSTRACT*

*This paper presents a security paradigm for edge devices to defend against various internal and external threats. The first section of the manuscript proposes employing machine learning models to identify MQTT-based (Message Queue Telemetry Transport) attacks using the Intrusion Detection and Prevention System (IDPS) for edge nodes. Because the Machine Learning (ML) model cannot be trained directly on low-performance platforms (such as edge devices), a new methodology for updating ML models is proposed to provide a tradeoff between the model performance and the computational complexity. The proposed methodology involves training the model on a high-performance computing platform and then installing the trained model as a detection engine on low-performance platforms (such as the edge node of the edge layer) to identify new attacks. Multiple security techniques have been employed in the second half of the manuscript to verify that the exchanged trained model and the exchanged data files are valid and undiscoverable (information authenticity and privacy) and that the source (such as a fog node or edge device) is indeed what it it claimed to be (source authentication and message integrity). Finally, the proposed security paradigm is found to be effective against various internal and external threats and can be applied to a low-cost single-board computer (SBC).*

*Keywords:*

Anomaly Detection; MQTT packets; Intrusion Detection System (IDS); Firewall; Edge Device.

==============================================================================================

## 1. INTRODUCTION

In recent years, the Industrial Internet of Things (IIoT) has developed as the fastest-evolving revolutionary technology, with the capability of digitizing and connecting a wide range of industries, resulting in huge economic possibilities and worldwide GDP (Gross Domestic Product) growth. Smart homes, smart cities, smart grids, logistics, transportation, linked autos, and supply chain management are all examples of IIoT applications [1, 2]. Although the IIoT offers exciting possibilities for the development of many industrial applications, it is vulnerable to cyberattacks and requires stricter security standards [1].

The huge number of sensors, controllers, and actuators used in the IIoT network to make intelligent business choices generate a lot of information, which has caught the attention of hackers all over the world. IIoT-enabled devices and sensors, on the other hand, are deemed resource-constrained devices with restricted power, memory, and communication capabilities.[1, 2].

As a result, edge devices such as laptops, desktops, routers, mini servers, hand-held devices, and cellphones are utilized to connect sensors and cloud servers. These devices collect data from sensors and send it to local servers after doing any essential preprocessing. Nevertheless, since the number of IoT edge devices in the industry has grown rapidly, various security and privacy issues have surfaced, posing a significant threat to IIoT's security and trustworthiness. Intruders may be able to take advantage of these edge devices [2].

This might result in operational inefficiencies, as well as financial and reputational damage. One of the biggest issues in current industrial contexts is ensuring cybersecurity in the IIoT. It comprises virus protection, unwanted access prevention, and communication and physical privacy protection. By installing modern and comprehensive security procedures, the IIoT's security, privacy, and trustworthiness may be improved. An Intrusion Detection System (IDS) is one of the most important security keys in the context of IoT/IIoT security by monitoring networks for malicious activity or policy breaches [1-3].

IDS might be based on signatures or anomaly detection. Signature-based detection approaches that employ predetermined criteria are excellent at identifying assaults for previously observed patterns. Anomaly-based identification, on the other hand, is used to detect unknown assaults or attacks with a lack of clearly established patterns. Machine learning algorithms have recently been shown to prevent many security vulnerabilities and improve the performance of anomaly-based detection methods [4-6].

In this manuscript, a security model for edge devices in IIoT networks is presented to protect edge devices against various external and internal attacks. The main contributions are as follows:
- Developing a lightweight IDS for edge devices to identify MQTT-based threats using ML models.
- Proposing a new update strategy by retraining the ML model on rich-computing devices such as fog nodes, and then executing it on limited resources devices such as edge nodes to deal with new attacks.
- Multiple security techniques are used to guarantee that data exchanged between the fog node and edge device during the update strategy is reliable and not readily.

The outline of this manuscript is as follows: Section 2 demonstrates the related works and tabulates the comparison between them. The security issues for the IIoT network are declared in section 3. The description of the proposed security model for edge devices is confirmed in section 4. In section 5, the system evaluation and security assessment for the proposed security model are described.

## 2. RELATED WORKS

Various studies have yielded many approaches for IDSs in the workplace. Several machine learning-based IDS for IIoT network have been proposed in the literature. Table 1 reviews the state-of-the-art and related studies [1].

In 2019, this paper suggested the architecture and detection method. The bottom layer network is subjected to a novel machine learning-based intrusion detection method, which significantly improves detection accuracy without increasing training time. With the help of its great learning capability, a deep learning algorithm applied to the resource-rich upper layer network delivers improved detection accuracy. The hybrid architecture improves network security by combining the advantages of both machine learning and deep learning methods. Furthermore, hierarchical information interaction and resource allocation reduce the network's existing bandwidth and energy limit [7]. To discover anomalies in the water treatment process, this study [8] provides a supervised learning strategy that uses the Roll-forward technique for validation and Classification and Regression Trees (CART) with integrals for classification. It includes the ability to validate time-series data using Roll-forward validation, which is subsequently followed by classification using the CART with invariants. It is simulated with the Mininet tool, and the accuracies of the train and test are respectively 99.9% and 98.1%.

In 2020, the technique in this paper [9] is based on two linear feature extraction algorithms, namely Principal Component Analysis (PCA) and Linear Discriminant Analysis (LDA), and it monitors the actions of factory network traffic. To evaluate the packets of network connections from the UNSW-NB15 database and discover and report abnormalities like malevolent assaults, a Machine-Learning-based technique is utilized.

In 2021, this thesis [10] focuses on detecting and classifying DoS (Denial of Service) attacks in MQTT sensor networks. To acquire accurate data, a smart home scenario was constructed to present a realistic MQTT-based dataset that comprises valid and malicious flow-level traffic by collecting legitimate and harmful data. The thesis shows how to identify and categorize DoS assaults using online and offline machine learning technologies. The suggested online classifiers can detect assaults in a variety of flow durations, allowing researchers to see the trade-off between classifier performance and flow lengths. In the IIoT infrastructure, this article [11] offers an analytical investigation of identifying abnormalities, cyber-attacks, and malevolent actions in a cyber-physical of vital water infrastructure. The paper classifies anomalous occurrences using a variety of machine learning methods, including numerous assaults and IIoT hardware failures. For the study review of the suggested technique, a real-world dataset encompassing 15 anomalous instances of the regular system operation was investigated. The test scenarios included anything from hardware failure to sabotage of water SCADA (Supervisory Control and Data Acquisition) devices. CART and Naïve Bayes (NB) have the greatest precision, F1-score, accuracy, and recall values, according to the findings.

In 2022, this study [12] described an architecture for enabling predictive analytics at the edge of a production system while they are being generated on a cloud node. The goal of this research is to look into the usage of knowledge graphs to collect information from maintenance employees as well as the linkages between assets and sensors. The suggested framework was put to the test in a use case involving an aluminum manufacturing firm. The results demonstrated the potential for further examination of the mixture of machine learning and knowledge graphs, confirming the time savings for training ML models as well as applications that can provide a better vision to operators regarding the manufacturing process.

Although there are numbers of papers, like the paper in [13] which proposed a methodology that is somewhat similar to our proposed methodology, their proposed methodology was used to build and train a Convolution Neural Network (CNN) model based on images dataset (which is out the scope of our research) without focusing on any application (like Intrusion Detection System (IDS)) and/or applying on constrained resources devices (which is in the scope of our research).

## 3. SECURITY ISSUES OF IIOT NETWORK

When combining IIoT systems, multilevel architectures are widely employed. Figure 1 depicts a multi-tier IIoT structure in progress. The main three layers that make up the IIoT network topology are edge, fog, and cloud [14].

- The edge computing layer consists of billions of IIoT machines that are linked to edge devices. Actuators, sensors, security cameras, vehicles, smart machines, and smart home appliances are examples of IIoT equipment that have restricted resources. This layer uses edge devices to collect massive volumes of data from a variety of sources and applications, then sends it to higher levels [15, 16].
- The fog layer sits between the cloud layer and the edge layer. Each fog node is in charge of serving a certain part of the city. Each fog node communicates with the central server in the cloud layer regularly to ensure system resilience. Fog nodes are more powerful than edge nodes and are closer to the edge computing layer than the cloud layer [16].
- The cloud layer comprises high-performance servers and storage devices. The central management and database core are located on the cloud layer's servers. Furthermore, they are in charge of monitoring the fog layer and nodes' security, availability, activities, and services [17].

Table 1: The details for the recent research based on IDS for IoT/IIoT Networks.

| Ref. | Year | Dataset Used | Architecture Approach | Algorithm Utilized | Upgrading Policy | CIA Triad | Prevention and response activities | Validation Strategy |
|---|---|---|---|---|---|---|---|---|
| [8] | 2019 | SUTD | N/A | Roll-forward technique and CART | × | × | × | Simulation |
| [7] | 2019 | Two datasets | Hybrid | Machine learning and deep learning | √ | × | × | Simulation |
| [9] | 2020 | UNSW-NB15 | N/A | Binary classes and multi-classes | × | × | × | Simulation |
| [11] | 2021 | Real-word dataset | N/A | Various machine learning | × | × | × | Simulation |
| [10] | 2021 | Realistic MQTT-based dataset | Distributed | online and offline machine learning | √ | × | × | Simulation |
| [12] | 2022 | Custom dataset | Central | Two types of machine learning | √ | × | × | Simulation/ Emulation |

*Not available N/A*
*Confidentiality, Integrity, and Authentication Triad (CIA)*
*Singapore University of Technology and Design (SUTD).*

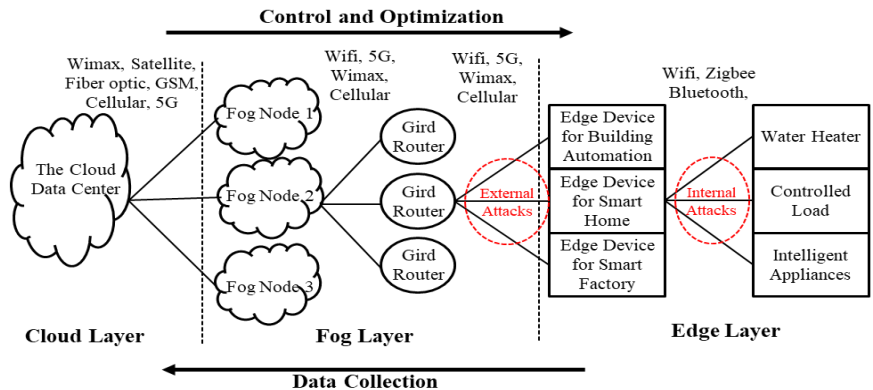

Fig. 1 An example of multi-tier IIoT architecture.

This manuscript focuses on the attacks on edge devices, which are essential components of the IIoT network. Security risks of edge devices can take many shapes and come from a variety of places. Insider attackers can be "machine users" (e.g., sensors) or faked edge devices. In other words, an insider attacker (attacks between the edge device and IIoT machines) is a network user who has some network expertise and uses it to study the architecture and configuration of the edge devices and the whole network. Outsider attackers (attacks between the edge device and fog node), on the other hand, exploit the Internet connection to launch assaults from a distant place outside of the coverage region. Edge devices are vulnerable to some assaults that vary in their nature, aims, and catastrophic results when researching the different sorts of attacks. An investigation of potential attacks against edge devices is conducted, based on their kind, with the descriptions summarized in Table 2.

## 4. DESCRIPTION OF THE PROPOSED SECURITY MODEL

The proposed security model is provided in this section to protect edge devices from various threats discussed before. The primary purpose is to safeguard edge device data (which is required for its functioning), as well as its accessibility and functionality.

Table 2: Investigation of the potential edge device attacks.

| Attacks Source | Attack Type | Description |
|---|---|---|
| Insider Attacks | Unauthorized IP address | It attempts to transfer data from an unauthorized device its IP (Internet Protocol) address is in the banned IP addresses. |
| | Unauthorized port No. | It attempts to transfer data using the destination port number found in the banned port list. |
| | Flooding Denial of Service | It's set up to keep the service from serving legitimate customers. |
| | Slow DoS against Internet of Things Environments (SlowITe) | It establishes many connections with the MQTT broker to simultaneously take advantage of all available connections. |
| | MQTT Publish Flood | It uses a single connection rather than numerous connections to saturate the resources. |
| | Malformed Data | Its goal is to generate and transmit multiple erroneous packets to the broker to cause exceptions on the specified service. |
| | Brute Force Authentication | Its goal is to crack clients' authentication credentials (password and username) during the authentication phase. |
| External Attacks | Sybil attack | It sends several messages to various edge devices, each with a forged source identity. |
| | Administrative Impersonation Attack | It claims the administrator's identity. |
| | Monitoring Attack | Passive monitoring of edge device activities. |
| | Edge Device Impersonation Attack | It claims the edge device identity. |
| | Data Sniffing and modification | It changes the data of the edge device or fog server and uses it for its benefit. |
| | Wireless Layer 2 attacks | It is disabling the network or compromising network users to glean sensitive information such as passwords. |

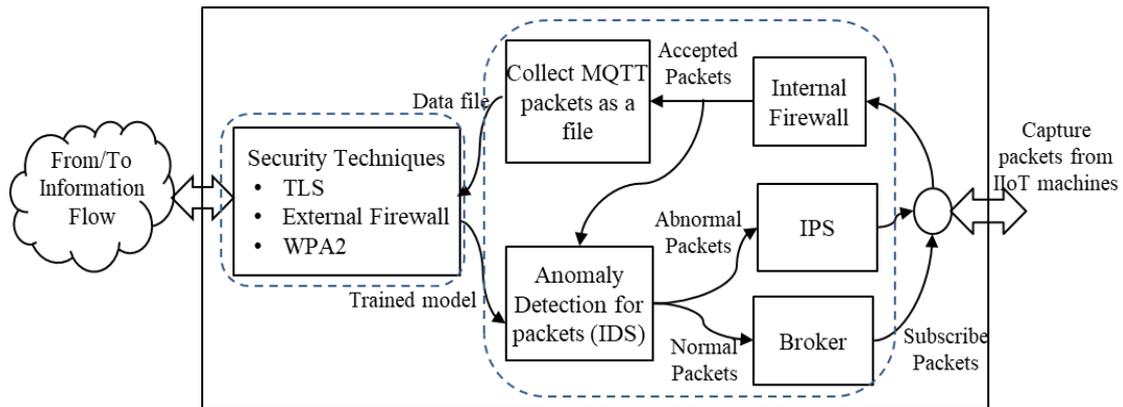

Fig. 2 The software structure of the edge device.

The proposed security model must meet several goals: it must ensure that the system is available and stable (using an Intrusion Detection System (IDS)), that the data exchanged is exact and untraceable (data privacy and authenticity), and that the source (e.g., fog server or edge device) is what it is claimed to be (source authentication and message integrity). Figure 2 shows the edge device's software architecture for implementing the recommended security paradigm. The recommended security paradigm is separated into two components, as shown in Figure 2. Each component is designed to protect each network port (assume the edge device has two different network connections). The proposed security model's first component protects the connection between edge devices and IIoT machines, while the second protects the connection between the fog server and edge device. The Raspberry Pi 4B platform may be utilized to build the suggested edge device since it is a compact, inexpensive, and independent single-board computer. The purpose of each role in the proposed edge node will be discussed in the following subsections.

### 4.1. Securing Connection Between Edge Device and IIoT Machines

The suggested edge device may connect wirelessly with IIoT equipment in the edge layer utilizing low-cost Wi-Fi-capable devices and the MQTT protocol as a communication mechanism. Because MQTT is a lightweight small-message protocol, it saves bandwidth and extends the battery life of the device. And since the MQTT client does not require a request update, the Publish / Subscribe architecture is more suited to the edge layer than the Request / Response architecture utilized in other protocols [18]. As a result, different threats such as DoS, brute force attacks, and others are launched against these types of packets, which may be detected by inspecting the header information in each MQTT packet. So, this part of the proposed security model has numerous roles, as follows:

- MQTT Broker: To serve as an MQTT broker, Mosquitto must be installed on the edge device [19]. Mosquitto is an open-source MQTT broker which offers a lightweight MQTT protocol server implementation.
- Internal Firewall: It is used to create a trusted wall of restrictions between IIoT machines and the MQTT broker by analyzing the network traffic using the IP and port addresses to see if it matches a set of rules defining what data flows are allowed to pass through it. The employed firewall blocks packets if their source IP addresses or specified prefix IP or destination port addresses are in the list of banned addresses and blocks too many requests made by the same IP quickly.
- IPS (Intrusion Prevention System): It is used to take an action automatically to stop or neutralize a potential attack when it is detected. The actions for the IPS module are: resetting the connection, forwarding the packet, and dropping the packet.
- IDS (Intrusion Detection System): It is a multi-class machine learning trained model which is used to constrain the normality for the packet. If the packet is normal, it will be directed to the MQTT broker. Otherwise, the attack's type will be detected, and the IPS module will be activated.

After the packet has been approved by the firewall, packet features are taken from it and are preprocessed to match the input of the trained model. The features are then recorded in a file that is sent to the fog server every so often (one week, one month, or whatever) to retrain the intrusion detection model. Figure 3 depicts the IDS module's flowchart for MQTT packets.

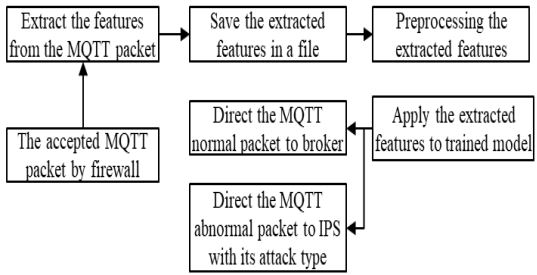

Fig. 3 The processing of the Intrusion detection for MQTT packets.

The intrusion detection model is a mixture of numerous processes. It is implemented and validated to classify multiple threats. To evaluate this scheme, a dataset must be selected. An open-source dataset named MQTTset was obtained from Kaggle [20]. This dataset was provided by Ivan Vaccari et al. [21]. In the MQTTset dataset, six types of labels represent one type of legitimate activities (Legitimate) and five types of threats, respectively (malformed, DoS, brute force, slowite, and flood), as shown in Table 2. The performance of different ML models is studied in this subsection to accurately detect the assaults and anomalies on packets. Random Forest (RF), Decision Tree (DT), and Gradient Boosting (GB) are the machine learning algorithms employed here. For hyper-parameters tuning, the best splitter, the Gini criterion, and the greatest depth were utilized in the DT until all leaves were genuine, while the RF was examined using two extreme estimators. The GB, on the other hand, is limited to a maximum of 20 estimators.

The used dataset is divided into 70–30 ratio into training and testing sets. The training set is then split into two subsets: A training set and a validation set, with a split ratio of 0.33 %, i.e., 67 % of the training set would be utilized for training and 33 % for validating. Only 18 features out of 33 in the dataset played an important role in the decision because the rest had a feature importance value of 0. On the new features set, the ML models for identifying attack packets have been trained and evaluated. To identify which technique is best suited for the proposed system, the models' performance is evaluated using key performance metrics such as Detection Rate, Precision, F-score, Recall, and Area Under the Curve (AUC), as well as Accuracy (as shown in Table 3). From Table 3, it can be noticed that DT and RF have higher accuracy, precision, recall, and F1 score values than GB. On the other hand, DT and RF are slightly more accurate than GB because GB may not be a good choice if the used dataset (MQTTset dataset) has a lot of noise. The AUC values for RF, DT, and GB are depicted in Table 4.

Table 3: The different evaluation metrics for the MQTTset dataset based on different classifiers.

|  | Metrics | RF | DT | GB |
|---|---|---|---|---|
| Training | Accuracy | 0.9969 | 0.9970 | 0.9912 |
|  | Precision | 0.9968 | 0.9969 | 0.9936 |
|  | Recall | 0.9968 | 0.9969 | 0.9912 |
|  | F1 Score | 0.9966 | 0.9967 | 0.9911 |
| Testing | Accuracy | 0.9968 | 0.9968 | 0.9903 |
|  | Precision | 0.9967 | 0.9967 | 0.9934 |
|  | Recall | 0.9968 | 0.9968 | 0.9903 |
|  | F1 Score | 0.9966 | 0.9966 | 0.9900 |

Table 4: The AUC values for different classifiers.

| Attacks Type | RF | DT | GB |
|---|---|---|---|
| Malformed | 0.96 | 0.95 | 0.58 |
| DOS | 0.99 | 0.99 | 0.45 |
| Flood | 0.96 | 0.95 | 0.91 |
| Legitimate | 0.99 | 0.99 | 0.98 |
| Brute Force | 0.94 | 0.93 | 0.86 |
| SlowITe | 0.97 | 0.97 | 0.93 |

According to the Area Under the Curves (AUC) in Table 4, DT and RF have higher accuracy because they achieve a 99% detection rate in some attacks. The AUC values for some classes are roughly equal to value one. In the case of GB, the AUC value is nearly equal to one for the legitimate class only, and the detection rates for some classes are very low, while other classes have close to 90% detection rates.

Table 5: The important times for the MQTTset dataset based on different classifiers.

| Algorithms | Train Time(sec) | Test Time(sec) | Model Size(KB) | Prediction Time(msec) |
|---|---|---|---|---|
| RF | 22.473942 | 2.1602210 | 742 | 1.67393 |
| DT | 33.585477 | 1.7004511 | 497 | 0.66304 |
| GB | 982.01197 | 7.3737800 | 137 | 1.51801 |

From Table 5, it can be distinguished that RF takes the minimum train time, but GB takes the maximum train time because RF builds multiple decision trees independently at the same time while GB builds multiple decision trees serially (one tree at each time). In the testing state, however, DT takes the lowermost time. It can also be noticed that the DT model consumes the least time when executed on the SBC (Raspberry Pi4). Finally, based on the used platform to build the edge device, the time consumed to extract the features from real packets for different trained models is 0.505978 msec. In conclusion, the proposed edge device can be based on the DT to detect the abnormality in the MQTT packets.

### 4.2. Securing Connection Between Edge Device and Fog Node

As previously stated, the proposed security model is also dependent on the fog node and edge devices interaction. The remote intrusion detection retrained technique performed by the fog node is a dangerous and sensitive activity that must be carried out with extreme caution to ensure the proper operation of edge devices. The following approach is proposed in this manuscript for maintaining the security of the intrusion detection model transfer:

Initially, a fresh data file is collected on the edge device, which is a single-board computer (SBC) (Raspberry Pi 4) provided with an anomaly detection model and security techniques tools. Nevertheless, to guarantee the freshness of the data file and to avoid replay assaults, a tag called the Data Version Tag (DVT) must be sent. The DVT is a crucial parameter. This tag contains the edge device's unique identifier, user identity, and date and time at creating this data file, and is used to determine the number of data files stored in the nearest fog node. The fog node also saves a copy of the edge device's unique identifier, the user identity in a trusted database. The DVT and data file are processed by security methods. Afterward, the handshaking is established to ensure that the fresh data file will be sent to the correct fog server. When the handshaking step is completed, the processed DVT and data file are ready to be sent over an unsecured network utilizing the TCP protocol (Transmission Control Protocol) for increased dependability. The fog server receives them through a wireless network port, processes them using inverted security mechanisms like decryption, and saves them in its memory. Later, the received data files will be used to rebuild and retrained the intrusion detection models.

Next, a new trained model file is created on the fog server based on the received data file from a certain edge device. The fog server is provided with machine learning and security techniques tools. Also, to ensure the freshness of the trained model and to avoid replay threats, another tag called the Trained Model Version Tag (TMVT) must be sent which contains the edge device's unique identifier, user identity, and date and time at creation of this trained model. This tag is used to determine the number of trained model updates by the fog server. The TMVT and trained model are sent to a certain edge device using the same steps mentioned earlier. Finally, the re-trained model will be exchanged with the old intrusion detection model. Figure 4 illustrates the transaction operations which are done between the fog server side and the edge device side.

*1) Transfer Data File and ML Model Based on CIA Triad*

Transport Layer Security (TLS) is a protocol used to provide Confidentiality, Integrity, and Authentication (CIA) between two communicating applications. It is used to ensure IoT/IIoT communication security and to protect connections between edge nodes and fog servers during the transfer of real datasets files and trained models over insecure channels [22, 23]. The trained model and the collected data file will be transferred in a separate TLS session.

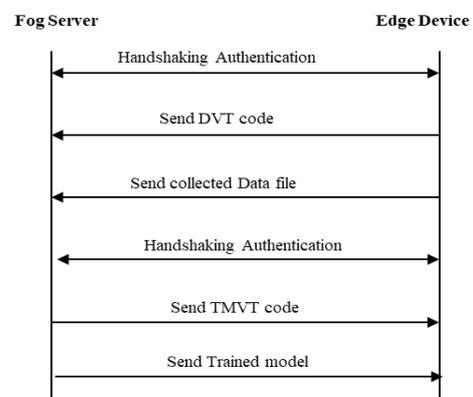

Fig. 4 The transactions between fog server-side and edge device side.

Table 6: Some assumptions to find latency between the edge node and fog server

| Parameter | value |
|---|---|
| Distance between the edge node and fog server | 100m |
| The Propagation Speed (PS) | $2 \times 10^8$ m/sec |
| Packet Processing Rate in Client (PPRC) (with effective for AES (Advanced Encryption Standard) and SHA (secure hashing algorithm) algorithms) | 10000 packet/sec |
| Packet Processing Rate in Server (PPRS) (with effective for AES and SHA algorithms) | 50000 packet/sec |
| SIFS | 20 μ sec. |
| DIFS | 300 μ sec. |
| Wireless Stander | 802.11a |
| Data Rate (DR) | 54 Mbps |
| The Size of Wireless Control signals (RTS, CTS, ACK, FIN) (SWC) | 74 bytes |
| TCP Header(TCPH) | 20 bytes |
| IP Header (IPH) | 20 bytes |
| Wireless Header (WH) | 34 bytes |
| The size of control TCP signals (REQ, ACK, GET, FIN) in the wireless network (SCTCP) | 74 bytes |
| The maximum Packet length in a Wireless network (MPLW) | 2346 bytes |

The communication time to transfer one trained model will be calculated in a theory that is based on some assumptions as shown in Table 6 [24]. Figure 5 shows the wireless and TCP signals to transfer a file (which can be a trained model file) from the FTP (File Transfer Protocol) server (Fog server) to the FTP client (edge node).

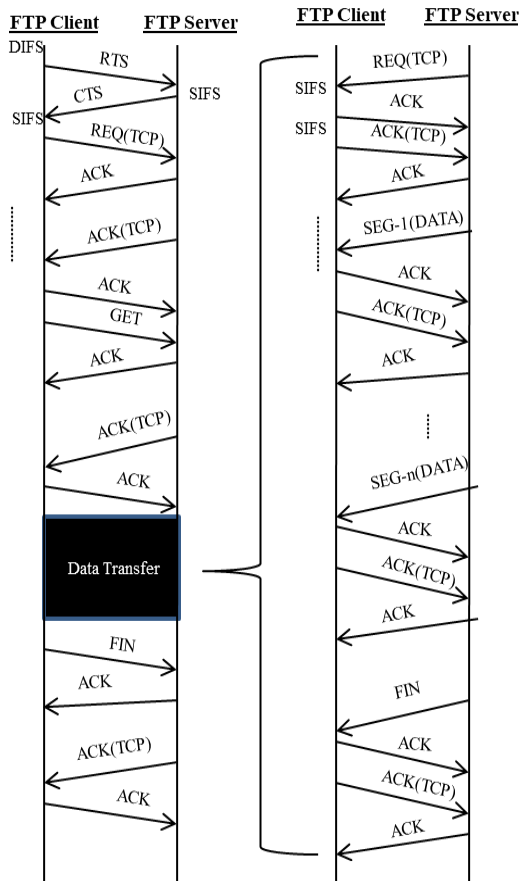

Fig. 5 The wireless and TCP signals to transfer a file from the FTP server to the FTP client

Some notes based on Figure 5 are stated as follows:
- Wireless signals (RTS (Request to Send), CTS (Clear to Send), and ACK (Acknowledgment)) have passed through Layer 2. As a result, the packet processing rate for them is not computed.
- The data rate for wireless signals is always (1 Mbps) when the system uses the 802.11x standard.
- The TCP control signals are REQ (TCP), ACK (TCP), and FIN (TCP). They are handled like data packets, which means they will require an ACK Wireless signal.
- When determining the latency of TCP control signals, the packet processing rate is taken into consideration.

To find the latency, it can be used the following equations [24]:

$$TCP\ Latency = Node\ delay + N.W\ delay \dots (1)$$

$$Node\ Delay = (No.\ of\ data\ packets + No.\ of\ TCP\ control\ Signal) * \left(\frac{1}{PPRC} + \frac{1}{PPRS}\right) \dots (2)$$

No. of data packets (N) required to transfer the file can be computed using:

$$N = \frac{FS * 8}{(MPLW - WH - IPH - TCPH) * 8} \dots (3)$$

$$Packet\ Delay = \frac{Packet\ Length * 8}{DR} + \frac{Distance}{PS} \dots (4)$$

TLS RSA WITH AES 128 CBC SHA is the cipher suite used in this section, which uses RSA (Rivest–Shamir–Adleman) for key exchange, AES-128 in Cipher Block Chaining (CBC) mode for encryption, and HMAC SHA1 for message authentication. Based on the TLS protocol version (1.2) and ciphersuite (TLS_RSA_WITH_AES_128_CBC_SHA), Table 7 [22, 23] displays the assumptions for the sizes of each message to find TLS overhead. On average, the overall overhead time to start a new TLS session is around 2.65msec. Figure 6 shows the latency to transfer different file sizes using the TLS protocol.

Table 7: The assumptions to find TLS overhead

| Parameter | Value |
| --- | --- |
| The average size of the initial ClientHello | 170 bytes |
| The average size of ServerHello | 75 bytes |
| The average size per Certificate | 1500 bytes |
| No. Of certificate in the chain | 4 |
| The average size of ClientKeyExchange for RSA server certificate. | 130 bytes |
| The average size of Finish message | 12 bytes |
| TLS Handshake header | 4 bytes |
| TLS Record header for each record sent | 5 bytes |

2) *External Firewall*

The traffic between the fog server and edge nodes and vice versa is monitored using the firewall which is a security device used to stop or mitigate unauthorized access to the edge device. The proposed system implements a software firewall to create a trusted wall of restrictions between the fog server and the edge device. It analyzes the network traffic using the IP and port addresses to see if they match a set of rules defining what data flows are allowed to pass through the firewall. The firewall is implemented using the NetfilterQueue library. The applied firewall performed the following functions:

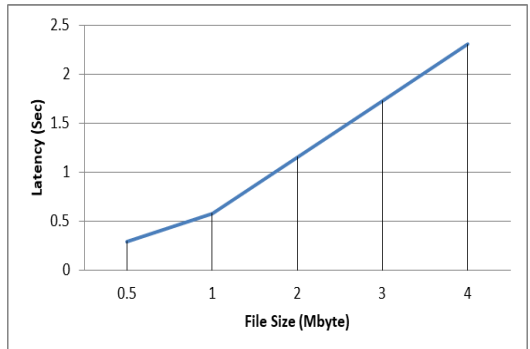

Fig. 6 The latency for different file sizes using TLS Protocol

- Block packets if their source IP address is in the list of banned IP addresses.
- Block packets if their destination port address is in the banned port addresses list.
- Block subnets by blocking specified prefixes of IP addresses found in the list of banned prefix addresses.
- Block too many requests made by the same IP quickly, such as ping attacks, by specifying the packet threshold and time threshold.

*3) Secure Wireless Communication*

Encrypting wireless data is crucial to help maintain a level of security because Wi-Fi signals are carried via the air and can be accessed by someone nearby. The security protocol known as Wifi Protected Access II (WPA2) was created to protect 802.11 wireless networks by encrypting sensitive data using 128-bit technology and encrypting passwords. Since 2006, WPA2, which is based on the IEEE 802.11i technical standard for data encryption, has been deployed on all certified Wifi equipment.

Temporal Key Integrity Protocol (TKIP) and Advanced Encryption Standard (AES) are the two encryption options often available to users when configuring a WPA2 network [25].

To produce certain encryption keys for the WPA2 protocol that is used to encrypt actual data delivered over wireless media, the authenticator (access point) and the applicant (client device) exchange 4 messages during the 4-way handshake procedure. These keys are produced via a four-way handshake and come from some source key material [26].

Using WPA2 decreases the performance of network connections due to the extra processing load of encryption and decryption. The performance impact of WPA2 is usually negligible, especially when compared with the increased security risk of using WPA or Wired Equivalent Privacy (WEP), or no encryption at all.

## 5. SYSTEM EVALUATION AND SECURITY ASSESSMENT

Different assessment criteria will be utilized to analyze the overall performance of the recommended edge device in this section. The proposed system's resource utilization statistics and network performance are listed in Table 8.

The primary takeaway from the statistics on system resource utilization shows that the proposed IDS was effectively and efficiently incorporated into the edge device platform. The proposed IDS utilizes a suitable amount of system resources while having minimal impact on the edge device's original functions. This architecture, on the other hand, assures that the addition of new IDS duties will not significantly decrease the node or network performance.

To round out the vision, a detailed security assessment for internal and external threats was conducted, taking into account the most likely attack vectors and sources of risk, and recommending effective remedies. According to Table 9, the suggested IDS can detect and protect against internal assaults that are MQTT-based threats. The proposed security methods are intended to defend against external threats, as shown in Table 10. Finally, the suggested edge device has a wealth of security features as well as realistic resource consumption.

Table 8: The proposed system's resource utilization statistics and network performance.

|  | Parameters | Deactivated IDS | Activated IDS |
|---|---|---|---|
| Network Performance | Average Round Trip Time (RTT) | 0.1544513 sec | 0.2622866 sec |
|  | Average No. of retransmission synchronous packets | 2 packets | 4 packets |
|  | Average No. of retransmission data packets | 5 packets | 13 packets |
|  | Average Throughput | 5786 bit/sec | 4662.7 bit/sec |
| Resource Utilization | Total Memory Utilization% | 1.73%(0.139 Gbyte) | 2.52%(0.202 Gbyte) |
|  | Average CPU Operation % | 5% | 16% |
|  | Average power Consumption | 3.325 W | 3.750 W |

Table 9: The security assessments for the internal attacks.

| Attack Type | Detected by | Response |
|---|---|---|
| Unauthorized IP address | Firewall | It is blocked by firewall |
| Unauthorized port No. | Firewall | It is blocked by firewall |
| Abnormal periodic data meter | Anomaly Detection for Metering Data | The user is informed by writing the details on the web page |
| Abnormal daily data meter | Anomaly Detection for Metering Data | The user is informed by writing the details on the web page |
| Flooding Denial of Service | IDS for Packet-level module | It is dropped by IPS |
| Slow DoS against Internet of Things Environments (SlowITe) | IDS for Packet-level module | IPS is sent the reset connection |
| MQTT Publish Flood | IDS for Packet-level module | IPS is forward to the same source IP address |
| Malformed Data | IDS for Packet-level module | It is dropped by IPS |
| Brute Force Authentication | IDS for Packet-level module | IPS is sent the reset connection |

Table 10: The security assessment for external attacks.

| Attack Type | Attacks' Target | Defense Strategy |
|---|---|---|
| Sybil attack | Edge Device services | Handshaking Authentication |
| Administrative Impersonation Attack | ▪Edge Device data & services<br>▪Fog server Functionality | Handshaking and Message Authentication |
| Monitoring Attack | Edge Device data & services | Packet Encryption |
| Illegal Access Attack | Edge Device data & services | Handshaking Authentication |
| Edge Device Impersonation Attack | ▪Edge Device data & services<br>▪Fog server Functionality | Malicious Edge Devices Detection |
| Data Sniffing and modification | ▪Edge Device data & services<br>▪Fog server Functionality | Message Authentication and Integrity |
| Wireless Layer 2 Attack | ▪Disable the network<br>▪ Compromise the network users | WPA2 |

## 6. CONCLUSION

Since edge devices connect IIoT machines to the data center, they are critical to IIoT networking. As a result, it is vital to safeguard these devices from numerous risks. The manuscript's first section proposed a security paradigm for edge devices to defend against various internal and external attacks. By leveraging machine learning methods, MQTT-based threats are identified, utilizing Intrusion Detection and Prevention SysteIDPS)-based security form (edge nodes). After the performance comparison between three ML algorithms (DT, RF, GB), The proposed edge device can be based on the DT to detect the abnormality in the MQTT packets because of its low prediction time (0.66304msec) and good accuracy with the MQTTset dataset (99.69%). Because the machine-learning model cannot be trained directly on low-performance devices (like edge devices), a new methodology for updating ML models is proposed, which involves training the model on a high-performance computing platform (such as the fog node) and then installing the trained model as a detection engine on low-performance platforms (such as the edge node of the edge layer) to detect new attacks. In the second half of the manuscript, the CIA triad, firewall, and WPA2 techniques are utilized to provide the major security aspects for the data transmitted (such as the trained model and gathered data files). A low-cost single-board computer (SBC) like the Raspberry Pi may easily implement the proposed security paradigm for edge devices. The security structure of the proposed edge device is lightweight (which has 2.52.% total memory utilization and 16% average CPU operation) and saves power (which has 3.75W average power consumption). As this paradigm is effective against a variety of internal and external assaults, it provides a balance between good performance and high security. Finally, it is hoped that this work will serve as a springboard for future advancements in cybersecurity analytics on edge devices employing effective machine learning models.

# نموذج أمان فعال لنظام إنترنت الأشياء الصناعي (IIoT) استنادًا إلى مبادئ التعلم الآلي

سحر لازم قدوري *                             قتيبة ابراهيم علي**
sahar.qaddoori@uoninevah.edu.iq                    qut1974@gmail.com

*جامعة نينوى - كلية هندسة الالكترونيات - قسم هندسة الالكترونيك- موصل – العراق
** جامعة الموصل - كلية الهندسة - قسم هندسة الحاسوب - موصل – العراق

**الملخص**

اقترحت هذه الورقة نموذجًا أمنيًا للأجهزة الحافة للدفاع عنها ضد التهديدات الداخلية والخارجية المختلفة. اقترح القسم الأول من المخطوطة استخدام نماذج التعلم الآلي لتحديد الهجمات المستندة إلى MQTT (نقل الرسائل عن بُعد في قائمة انتظار الرسائل) باستخدام نظام كشف ومنع التطفل (IDPS) للعقد الطرفية. نظرًا لأنه لا يمكن تدريب نموذج التعلم الآلي مباشرة على منصات منخفضة الأداء (مثل الأجهزة الحافة) ، لذلك ، تم اقتراح منهجية جديدة لتحديث نماذج ML ولتوفير مفاضلة بين أداء النموذج والتعقيد الحسابي . تضمنت المنهجية المقترحة تدريب النموذج على منصة حوسبة عالية الأداء ثم تثبيت النموذج المدرب كمحرك كشف على منصات منخفضة الأداء (مثل عقدة الحافة لطبقة الحافة) لتحديد الهجمات الجديدة. تم استخدام تقنيات أمان متعددة في النصف الثاني من المخطوطة للتحقق من أن النموذج المدرب المتبادل وملفات البيانات المتبادلة صالحة وغير قابلة للاكتشاف (صحة المعلومات والخصوصية) وأن المصدر (مثل عقدة الضباب أو جهاز الحافة) هو في الواقع ما تدعي أنه (مصادقة المصدر وسلامة الرسالة). أخيرًا ، يعتبر نموذج الأمان المقترح فعالًا ضد التهديدات الداخلية والخارجية المختلفة ويمكن تطبيقه على كمبيوتر منخفض التكلفة أحادي اللوحة (SBC).

**الكلمات الداله :**

الكشف الشاذ، حزم ال MQTT، نظام كشف التطفل، الجدار الناري، امنية طبقة النقل (TLS)، جهاز الحافة.